\documentclass[twocolumn, showpacs, prl]{revtex4-1}
\usepackage{graphicx, color}

\def\ind#1{{_{\mathrm{#1}}}}

\begin{document}

\title{Creation and Amplification of Electro-Magnon Solitons by Electric Field in Nanostructured Multiferroics}
\author{R. Khomeriki$^{1,2}$, L. Chotorlishvili$^{1}$, B. A. Malomed$^{3}$,
J. Berakdar$^{1}$}

\address{$^1$Institut f\"ur Physik, Martin-Luther Universit\"at Halle-Wittenberg, D-06120 Halle/Saale, Germany \\
$^2$Physics Department, Tbilisi State University, 0128 Tbilisi, Georgia \\
$^3$Department of Interdisciplinary Studies, School of Electrical
Engineering, Faculty of Engineering, Tel Aviv University, Tel Aviv
69978, Israel}

\begin{abstract}
We develop a theoretical description of electro-magnon solitons in a coupled
ferroelectric-ferromagnetic  heterostructure. The solitons are considered  in the weakly
nonlinear limit as a modulation of plane waves corresponding to two,
electric- and magnetic-like branches in the spectrum.
Emphasis is put  on magnetic-like envelope solitons that can be created by
an alternating electric field. It is shown also that the magnetic pulses
can be amplified by an electric field with a frequency  close to the band
edge of the magnetic branch.
\end{abstract}

\pacs{85.80.Jm, 75.78.-n, 77.80.Fm}
\maketitle

\address{$^1$Institut f\"ur Physik, Martin-Luther Universit\"at Halle-Wittenberg, D-06120 Halle/Saale, Germany \\
$^2$Physics Department, Tbilisi State University, 0128 Tbilisi, Georgia \\
$^3$Department of Interdisciplinary Studies, School of Electrical
Engineering, Faculty of Engineering, Tel Aviv University, Tel Aviv
69978, Israel}

Multiferroic materials, i.e., materials exhibiting coupled order parameters,
are in the focus of current research. These systems offer not only new
opportunities for applications but also provide a
test ground for addressing fundamental issues regarding the interplay between
electronic correlations, symmetry, and the interrelation between magnetism and ferroelectricity \cite{ME-review,Single-ME,Composite-ME}. Here we address magnetoelectrics which possess
a simultaneous ferroelectric-magnetic response.
A interesting  aspect is the non-linear nature of the magnetoelectric  excitation dynamics, which hints at
the potential of these systems for exploring  nonlinear
wave-localization phenomena, such as multicomponent solitons
\cite{malomed1,malomed2}, nonlinear band-gap transmission
\cite{leon,ram1}, and the interplay between the nonlinearity and Anderson
localization \cite{flach}.
 In this paper we aim at exciting robust magnetic
signals by means of electric fields. Particularly, we consider
a multiferroic nano-heterostructure consisting of a ferromagnetic (FM) part deposited onto
 a ferroelectric (FE) substrate. As demonstrated experimentally, under favorable conditions,
 a coupling between the ferroelectric and the ferromagnetic order parameters may emerge (this coupling is referred to as
 the magnetoelectric coupling), thus allowing one to control magnetism (ferroelectricity) by means of electric (magnetic) fields.
 Here we consider the case when the multiferroic structure is driven by an electric field with a frequency
  located within the band-gap of the FE branch and in
the band of the magnetic-excitation branch. For a proper choice
of the electric-field frequency (that follows from the
electro-magnon soliton theory developed below) it is possible to excite propagating magnetic
solitons.
 In addition, we point out a possibility for the amplification of weak magnetic
signals, which suggests the design of a digital magnetic transistor,
where the role of the pump is played by the electric field.

Examples of the two-phase multiferroics under study \cite{BoRu76,SpFi05,Fi05,levan1},
are BaTiO$_{3}$/CoFe$_{2}$O$_{4}$ or PbZr${_{\mathrm{1-x}}}$Ti${_{%
\mathrm{x}}}$O${_{\mathrm{3}}}$/ferrites. The developed model will be applied
 to a system where the FE  and FM regions
 are coupled at an interface whith a weak
magnetoelectric coupling. The theory is, however,
 more general and can, in principle, be applied
to single-phase magnetoelectrics \cite{RaSp07,Dz59,levan2}. For the creation of electro-magnon solitons,
which is the subject of the present work,  a two-phase multiferroic structure
is  more appropriate, as it allows  to generate and
manipulate isolated FE or FM signals away from the
interface.

Both single- and two-phase multiferroics may be modeled by a ladder
consisting of two weakly coupled chains: One chain is ferroelectric
(FE)  built out of unidimensional electric dipole moments, $P_{n}$. The
second chain is  ferromagnetic (FM), composed of classical
three-dimensional magnetic moments, ${\vec{S}_{n}}$, where $n$ numbers
the site  in the lattice. Each chain is characterized by an
intrinsic nearest-neighbor coupling, and each  $P_n$ is coupled to
 $\vec S_n$ via inter-chain weak magnetoelectric coupling. For a
discussion of the microscopic nature of this coupling we refer to
~\cite{jam}. We assume the direction of FE dipoles at some arbitrary
angle with respect to FM anisotropy axis $\xi$, as depicted in Fig. 1a.
The magnetoelectric coupling will cause a rearrangement of magnetic moments.
 Let the new ground-state ordering direction of FM be the
axis $z$, and $\phi$ is the angle between $z$ and anisotropy axis
$\xi$. The magnetic field $\vec h(t)$ is applied along $z$, and
$\theta$ is an angle between $z$ and FE moments
(see Fig. 1a). $S_0$ ($P_0$) stands for the FM (FE) equilibrium configuration.
 We will consider
perturbations around the equilibrium. Defining the  scaled
dipolar deviations $p_n\equiv(P_n-P_0)/P_0$ and the scaled magnetic
variables $\vec s_n\equiv \vec S_n/S_0$, the Hamiltonian is written as
\begin{eqnarray}
&&H = H\ind{P}  + H\ind{S} + H\ind{SP}, \qquad H\ind{SP} = -\tilde g
\sum\limits_{n =1}^N p_n  s_n^x,  \label{1}\\
&&H\ind{P}  = \sum\limits_{n = 1}^N { \frac{\tilde\alpha_0}{2}
\left(\frac{dp_n}{dt}\right)^2 + \frac{\tilde\alpha}{2}
p_n^2+\frac{\tilde\beta}{4}p_n^4+ \frac{\tilde\alpha_J}{2}\left(
{p_{n + 1} - p_n } \right)^2 },
\nonumber \\
&&H\ind{S}  = \sum\limits_{n =1}^N \left[ -\tilde J\vec s_n \vec
s_{n + 1} + \tilde D_1\left( {s_n^x } \right)^2+ \tilde D_2\left(
{s_n^y } \right)^2\right] , \nonumber
\end{eqnarray}
where $H\ind{SP}$ stands for the linearized interfacial magnetoelectric coupling
 between the FM
and the FE chain \cite{jam}.
 $H\ind{P}$ is FE part of the
energy functional for  $N$-interacting FE dipole moments
\cite{SuJi10,GiChGu11}. Further, $\tilde\alpha_0$ is a kinetic coefficient;
$\tilde\alpha_J$ is the nearest-neighbor coupling constant;
$\tilde\alpha$ and $\tilde\beta$ are second- and forth-order
expansion coefficients of the Ginzburg-Landau-Devonshire (GLD)
potential \cite{RaAh07,SuJi10} near the equilibrium state $P_0$.
$H\ind{S}$ stands the ferromagnetic contribution ~\cite{Ch02}, where $J$
is the nearest-neighbor exchange coupling in the FM part. $\tilde
D_1=\tilde D S_0\cos^2\phi$ and $\tilde D_2=\tilde D S_0$ are
anisotropy constants, and  $D$ is the uniaxial anisotropy constant
along axis $\xi$ (see Fig. 1b).
\begin{figure}[t]
\centering \includegraphics[scale=.47]{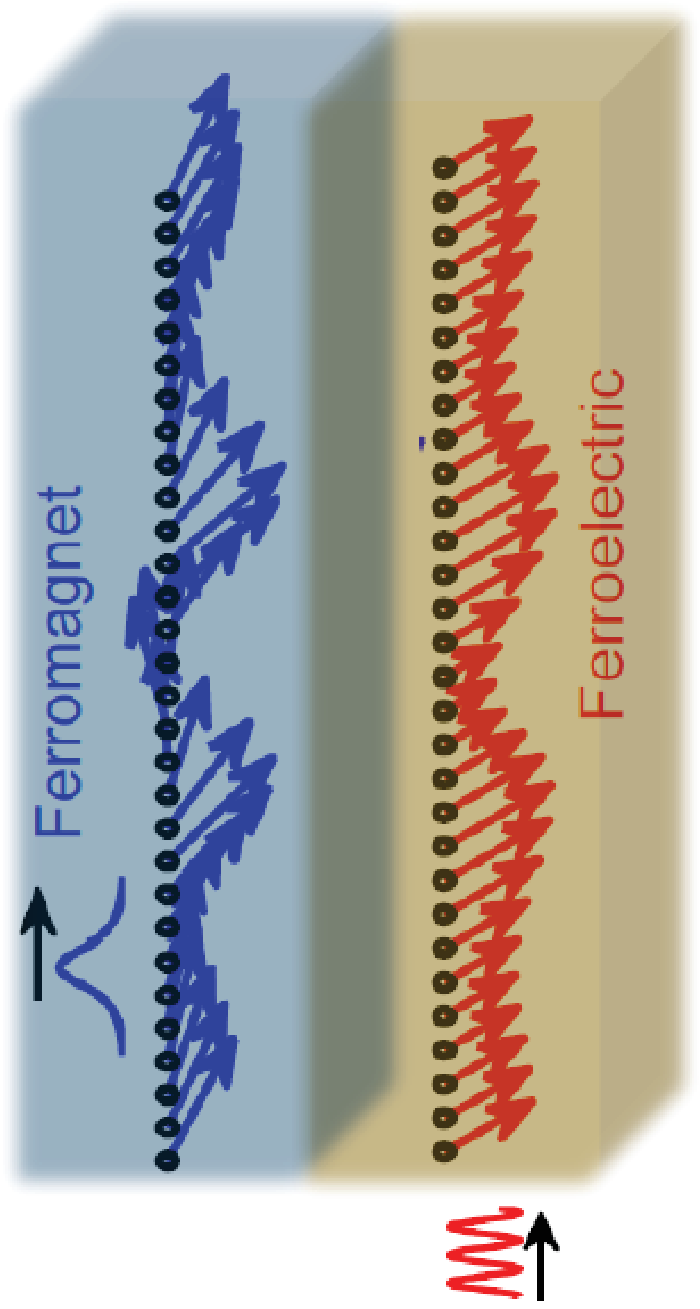}
\includegraphics[scale=.6]{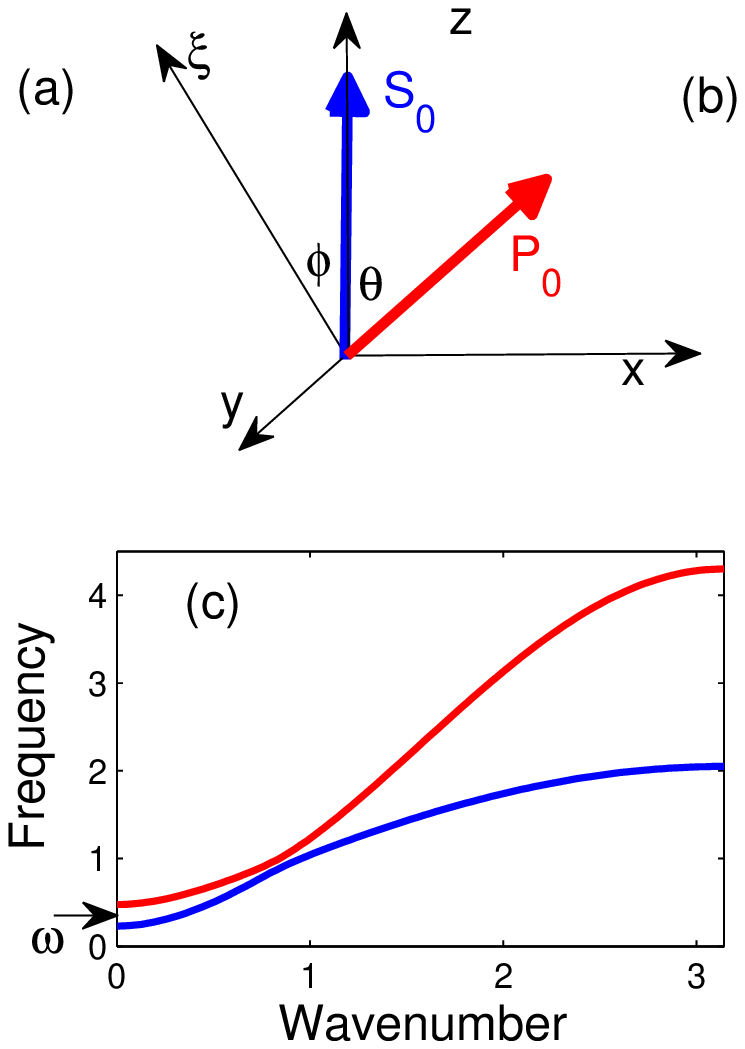} \caption{(Color
online) (a) A schematic of the multiferroic ladder built of FE and
FM chains. Arrows indicate the directions and the magnitudes of the
electric dipole moments and magnetic moments in the course of the
soliton advancement along the ladder. (b) Mutual orientations of FE
and FM ground-state vectors, in the framework of Hamiltonian
(\protect\ref{1}). (c) Dispersion relations for FE (upper) and FM
(lower) branches of the multiferroic composite. The arrow indicates
the selected carrier frequency.} \label{fig_0}
\end{figure}

We operate with dimensionless quantities by using the scaling
 $t\rightarrow\omega_0t$ with
$\omega_0=\sqrt{\alpha_J/\alpha_0}$ (for the examples shown below
$\omega_0\sim 10^{12}$ rad/sec). The other parameters of the model
$\tilde g$, $\tilde\alpha$, $\tilde\beta$, $\tilde J$, $\tilde D_1$,
$\tilde D_2$  are scaled with $\omega_0$. The scaled quantities are
indicated by omitting  the tilde superscript. The  time evolution is
governed  by
\begin{eqnarray}
&&\frac{\partial {s}_{n}^{x}}{\partial t}=-J\left[ s_{n}^{y}\left(
s_{n-1}^{z}+s_{n+1}^{z}\right) -s_{n}^{z}\left(
s_{n-1}^{y}+s_{n+1}^{y}\right) \right] -  \nonumber \\
&&-2D_{2}s_{n}^{y}s_{n}^{z},  \nonumber \\
&&\frac{\partial {s}_{n}^{y}}{\partial t}=J\left[ s_{n}^{x}\left(
s_{n-1}^{z}+s_{n+1}^{z}\right) -s_{n}^{z}\left(
s_{n-1}^{x}+s_{n+1}^{x}\right) \right] +  \nonumber \\
&&+2D_{1}s_{n}^{x}s_{n}^{z}-gp_{n}s_{n}^{z},  \label{3} \\
&&\frac{d^{2}p_{n}}{dt^{2}}=-\alpha p_{n}-\beta p_{n}^{3}+\left(
p_{n-1}-2p_{n}+p_{n+1}\right) +gs_{n}^{x}.  \nonumber
\end{eqnarray}%
As we are interested in small perturbations,
 $s_n^x$, $s_n^y$ and $p_n$ are much less than unity and
the approximate equality
$s_n^z=1-\left(s_n^x\right)^2/2-\left(s_n^y\right)^2/2$ is justified.

We seek weakly nonlinear harmonic solutions to
Eq. (\ref{3}), with a frequency $\omega $ and a wavenumber $k$, in the
form of a  column vector $\left( s_{n}^{x},s_{n}^{y},p_{n}\right)
=\mathbf{R}\exp \left[ i\left( \omega t-kn\right) \right]
+\mathrm{c.c.}$, where $\mathbf{R}$ is a set of complex amplitudes
$\mathbf{R}\equiv (a,b,c)$, and $\mathrm{c.c.}$\
stands for the complex conjugate. Neglecting higher harmonics in Eq. (\ref{3}),
 we find the set of nonlinear algebraic equations
\begin{equation}
\hat{\mathbf{W}}\ast \mathbf{R}=\mathbf{\emph{Q}}^{\mathrm{nl}},  \label{4}
\end{equation}%
where the matrix and source are, respectively,
\begin{widetext}
\[
\hat{\mathbf{W}}=\left(
\begin{array}{ccc}
i\omega  & J\sin ^{2}\left( k/2\right) +2D_{2} & 0 \\
-\left[ J\sin ^{2}\left( k/2\right) +2D_{2}\right]  & i\omega  & g \\
g & 0 & \omega ^{2}-\alpha -\sin ^{2}\left( k/2\right)
\end{array}%
\right) ,
\]%
\begin{eqnarray}
 \textbf{\emph{Q}}^{nl}= \left(
  \begin{array}{c}
    b(|b|^2+|a|^2)(J-J\cos k+D_2)-b^*(b^2+a^2)(J\cos k-J\cos 2k-D_2) \\
    -a(|b|^2+|a|^2)(J-J\cos k+D_1)+a^*(b^2+a^2)(J\cos k-J\cos 2k-D_1) \\
    -3\beta |c|^2c \\
  \end{array}
\right).
\end{eqnarray}
\end{widetext}

The linear limit amounts to the set of linear homogeneous algebraic
equations $\hat{\mathbf{W}}\ast \mathbf{R}=0$. The solvability
condition  $\mathrm{Det}\left( \hat{\mathbf{W}}\right) =0$
leads to two branches of the dispersion relation $\omega (k)$ which are
shown in Fig. 1(c), with the corresponding amplitude set,
$\mathbf{R}=(a,b,c)$, where $b$ and $c$ are expressed via the arbitrary
constant $a$: $b=-ia\omega /\left[ J\sin ^{2}(k/2)+D_{2}\right] $,
$c=ga/\left[ \alpha +\sin ^{2}(k/2)-\omega \right] $. We call a
dispersion branch ferroelectric (defining its frequency $\omega
_{E}$ and labeling the amplitude with index $E$), if it has
$|c|>|a|$ [the red curve in Fig. 1(c)], while a ferromagnetic branch
($\omega _{M}$) is defined by the relation $|c|<|a|$ (the blue curve in
the same figure).

Of a particular interest is the case when the system is excited at an edge
(at the left one, for the sake of definiteness), with a frequency
$\omega _{s}$ which falls into the bandgap of FE mode and,
simultaneously, the propagation band of the FM one, as shown by the
arrow in Fig. 1(c). The dispersion relation with the fixed
frequency, $\omega =\omega _{s}$ becomes then a cubic equation for $\sin
^{2}(k/2)$. For $\omega _{s}$
belonging to the band of FM mode and bandgap of FE one, the cubic
equation yields two complex wavenumbers, associated with FE and FM
modes, and a real one, corresponding to the FM mode. These three
wavenumbers determine a set of three orthogonal eigenvectors,
$\mathbf{R_{E}}\equiv (a_{E},b_{E},c_{E})$,
$\mathbf{R_{M}^{-}}\equiv (a_{M}^{-},b_{M}^{-},c_{M}^{-})$ and
$\mathbf{R_{M}^{+}\equiv (a_{M}^{+},b_{M}^{+},c_{M}^{+})}$, where
the first two correspond to complex FE and FM wavenumbers, $k_{E}$
and $k_{M}^{-}$, respectively, while the last one is related to the
real wavenumber, $k_{M}^{+}$. In linear systems, the solutions with
the complex wavenumbers are evanescent waves localized at the left edge
of the multiferroic chain. Thus, the solution for the vector
function, $\mathbf{F_{n}}=\left( s_{n}^{x},s_{n}^{y},p_{n}\right) $,
is
\begin{eqnarray}
\mathbf{F_{n}^{E}} &=&A(t)\mathbf{R_{E}}e^{i\omega
_{s}t-|k_{E}|n }+c.c.  \nonumber \\
\mathbf{F_{n}^{M^{-}}} &=&B(t)\mathbf{R_{M}^{-}}e^{i \omega
_{s}t-|k_{M}^{-}|n}+c.c.  \label{6}
\end{eqnarray}%
where the amplitudes $A(t)$ and $B(t)$ may  vary slowly in time.

As mentioned above, the solutions corresponding to the complex wavenumbers are
localized at the boundary. To examine the possibility of a solitonic
self-localization of the third solution with a real wavenumber (cf. Refs. \cite%
{new,kalinikos,ram} for similar solutions in multiferroic models),
we consider
the nonlinear frequency shift produced by the small terms $\mathbf{%
\emph{Q}}^{\mathrm{nl}}$ in (\ref{4}). Assuming a shifted frequency, $\omega
+\delta \omega $ instead of $\omega $, the matrix $\hat{\mathbf{W}}$ is
substituted by a modified one, $\hat{\mathbf{W}}+\mathbf{\delta \hat{W}}$,
with the diagonal matrix  $\mathbf{\delta \hat{W}}\equiv i\delta \omega
\cdot \mathrm{Diag}(1,1,2\omega )$. We also define a row vector $\mathbf{L}%
=(a^{\prime },b^{\prime },c^{\prime })$ which solves for the equation $\mathbf{L}%
\ast \hat{\mathbf{W}}=0$. Then, multiplying both sides of Eq. (\ref{4}) on $%
\mathbf{L}$, we obtain the nonlinear plane-wave frequency shift:
\begin{equation}
\delta \omega \left( k,|a|^{2}\right) =-i\left( \mathbf{L}\ast \mathbf{\emph{%
Q}}^{\mathrm{nl}}\right) \biggr/\left( \mathbf{L}\ast \mathbf{\delta \hat{W}%
\ast R}\right) .  \label{5}
\end{equation}

From these results, operating with the envelope function $\varphi(n,t)$
defined from $\mathbf{F_{n}^{M^{+}}}=\varphi(n,t)e^{i\left( \omega
_{s}t-ik_{M}^{+}n\right)}\left(a_{M}^{+},~b_{M}^{+},~c_{M}^{+}\right)$,
one can derive the nonlinear Schr\"{o}dinger equation (NLS), cf.
Refs. \cite{boardman,taniuti} in the form:
\begin{equation}
2i\left(\frac{\partial\varphi}{\partial
t}+v\frac{\partial\varphi}{\partial n}\right)+\omega ^{\prime \prime
}\frac{\partial^2\varphi}{\partial n^2}+\Delta|\varphi|^2\varphi=0,
\label{NLS0}
\end{equation}
which gives rise to the respective envelope-soliton solution with the
FM-like localized mode  being  written as
\begin{equation}
\mathbf{F_{n}^{M^{+}}}=\frac{e^{i\left( \omega _{s}t-k_{M}^{+}n\right) }}{{%
\cosh }\left[ \frac{a_{M}^{+}}{2}\sqrt{\frac{\Delta }{2\omega ^{\prime
\prime }}}(n-vt)\right] }\left(
\begin{array}{c}
a_{M}^{+} \\
b_{M}^{+} \\
c_{M}^{+} \\
\end{array}%
\right) \mathrm{+c.c.}.  \label{7}
\end{equation}%
The velocity, dispersion, and the nonlinearity coefficients are
\begin{equation}
v=\frac{\partial \omega _{M}}{\partial k},\quad \omega ^{\prime \prime }=-%
\frac{\partial ^{2}\omega _{M}}{\partial k^{2}},\quad \Delta =\frac{\partial %
\left[ \delta \omega _{M}(k,|a|^{2})\right] }{\partial \left[
|a|^{2}\right] }.
\end{equation}%
The carrier frequency of this soliton is defined by the dispersion
relation [the lower blue curve in Fig. 1(c)]:
\begin{equation}
\omega _{s}=\omega _{M}(k)+\delta \omega _{M}\left( k,|a|^{2}\right)
/2. \label{7m}
\end{equation}

Thus, one can generate both the FE evanescent (\ref{6}) and FM solitonic (%
\ref{7}) modes, driving the left edge of the chain at the same frequency, $%
\omega _{s}$. It is possible to produce a combination of these solutions
too. Generally in  nonlinear systems, linear combinations of particular
solutions is not another solution but if the solutions are far separated,
which makes interactions between them negligible, the linear combination
\begin{equation}
\mathbf{F_{n}}=f_{1}\mathbf{F_{n}^{E}}+f_{2}\mathbf{F_{n}^{M^{-}}}+\mathbf{%
F_{n}^{M^{+}}}  \label{8m}
\end{equation}%
is still a solution of the nonlinear problem. In the weakly nonlinear limit
it is even possible to construct a solution for the case when particular modes overlap
(i.e., the magnetic soliton is located near the edge), adding a
time-dependent phase to each term (\ref{6}) and (\ref{7}) in the sum \cite%
{oikawa}. For instance, one can consider an approximate solution at the left
edge of the ladder, $n=0$, in the form of
\begin{equation}
\mathbf{F_{0}}=f_{1}\mathbf{F_{0}^{E}}e^{i\Psi ^{E}(t)}+f_{2}\mathbf{%
F_{0}^{M^{-}}}e^{i\Psi ^{M^{-}}(t)}+\mathbf{F_{0}^{M^{+}}}e^{i\Psi
^{M^{+}}(t)},  \label{8}
\end{equation}%
where, in the weakly-nonlinear limit, the phases $\Psi $ are proportional to
the wave amplitudes. Hence the waves do not gain significant phase shifts due to
interaction effects, if their relative group velocity is not negligible. In
this case, all phase shifts may be neglected.
\begin{figure}[b]
\centering \includegraphics[scale=.52]{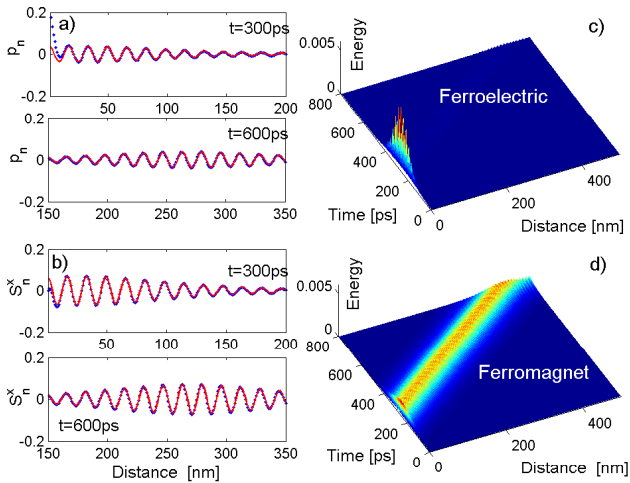}
\caption{(Color online) Numerical simulations of the creation of a
ferromagnetic soliton by the electric field, using Eq. (\protect\ref{3})
with parameters (\protect\ref{12}). (a,b) Show a comparison between the analytical
(solid lines) and the numerical (points) results for the soliton's spatial
profile at different moments of time. The analytical solution is taken
according to Eq. (\protect\ref{8m}) with the coefficients (\protect\ref{11}).
(c,d): The propagation of the FM soliton in the ferroelectric and the
ferromagnetic layers, respectively.}
\label{fig_1}
\end{figure}

Our particular aim is to create an FM soliton by exciting only the FE
degree of freedom at the edge, i.e., $s_{0}^{x}=s_{0}^{y}=0$. To this end,
we choose $A(t)=B(t)=\mathrm{sech}\left[ a_{M}^{+}vt\sqrt{\Delta \Bigr/%
8\omega ^{\prime \prime }}\right] $, seeking to impose the following vector
relation at the edge, $n=0$:
\begin{equation}
(0,0,p_{0})=\frac{\left( f_{1}\mathbf{R_{E}}+f_{2}\mathbf{R_{M}^{-}}+\mathbf{%
R_{M}^{+}}\right) e^{i\omega _{s}t}}{\cosh \left[ a_{M}^{+}vt\sqrt{\Delta
/8\omega ^{\prime \prime }}\right] }+\mathrm{c.c.}  \label{9}
\end{equation}

Using now the orthogonality of eigenvectors $\mathbf{R}$, we readily get the
appropriate expression for $p_{0}$:
\begin{equation}
p_{0}(t)=\frac{\Bigl|\mathbf{R_{M}^{+}}\Bigr|^{2}}{\left( c_{M}^{+}\right)
^{\ast }}\frac{e^{i\omega _{s}t}}{\cosh \left[ a_{M}^{+}vt\sqrt{\Delta
/8\omega ^{\prime \prime }}\right] }+\mathrm{c.c.}  \label{10}
\end{equation}%
Further, it is possible to compute the coefficients $f_{1}$ and $f_{2}$, using
the same orthogonality property:
\begin{equation}
f_{1}=\frac{\left\vert \mathbf{R_{M}^{+}}\right\vert ^{2}\left( c_{E}\right)
^{\ast }}{\left\vert \mathbf{R_{E}}\right\vert ^{2}\left( c_{M}^{+}\right)
^{\ast }};\quad f_{2}=\frac{\left\vert \mathbf{R_{M}^{+}}\right\vert
^{2}\left( c_{M}^{-}\right) ^{\ast }}{\left\vert \mathbf{R_{M}^{-}}%
\right\vert ^{2}\left( c_{M}^{+}\right) ^{\ast }}.  \label{11}
\end{equation}

In numerical simulations, if we make $p_{0}$ a function of time as per
Eq. (\ref{10}), keeping the magnetic moments pinned at the boundary, it is
possible to excite the FE and FM evanescent waves (\ref{6}), and also propagating
FM soliton (\ref{7}). These simulations correspond to an experimental setup
with pinned boundary conditions at both FE and FM edges, and to the application
of the electric field $E(t)=p_{0}(t)$ according to Eq. (\ref{10}) at the
first cell of the FE chain. In this way, one can realize the excitation of
magnetic solitons in the FM chain of the multiferroic ladder via an
electric (rather than magnetic) field by virtue of the magnetoelectric
coupling.

For an assessment of the above, we performed full numerical simulations
with  the following values of the normalized
parameters
\begin{equation}
\alpha =0.2;~\beta =0.1;~J=1;~D_{1}=0.1;~D_{2}=0.2;~g=0.1. \label{12}
\end{equation}%
These values  correspond to  $BaTiO_3/Fe$ \cite{DuJa06,LeSa10}.
 Furthermore, we assume for the FE second and fourth order potential
coefficients  $\tilde\alpha_1/(a^3_{\mathrm{FE}})=2.77\cdot 10^{7}$
[Vm/C], $\tilde\alpha_2/(a^3_{\mathrm{FE}})=1.7\cdot 10^8$
[Vm$^5$/C$^3$], and for the FE coupling coefficient
$\tilde\alpha_J/(a^3_{\mathrm{FE}})=1.3\cdot 10^8$ [Vm/C], the
equilibrium polarization $P_{0}=0.265$ [C/m$^2$], and the
coarse-grained FE cell size $a_{\mathrm{FE}}=1$ [nm]. The FM
exchange interaction strength is $\tilde J=3.15\cdot 10^{-20}$ [J],
the FM anisotropy constant is $\tilde D=6.75\cdot 10^{-21}$ [J], and
the ME coupling strength is $\tilde g_0\approx 10^{-21}$ [Vm$^2$].

We drive  the left boundary of FE according to
(\ref{10}) with a driving frequency $\omega_s=0.4$ and an amplitude
$a_M^+=0.07$ and apply  pinned boundary conditions for
FM, $s_{0}^{x}=s_{0}^{y}=0$. The results are displayed in
Fig. 2, where the comparison of numerical simulations and
approximate analytical solution (\ref{8m}) are shown  at different
moments of time [Figs. 2(a,b)].
\begin{figure}[t]
\centering \includegraphics[scale=.52]{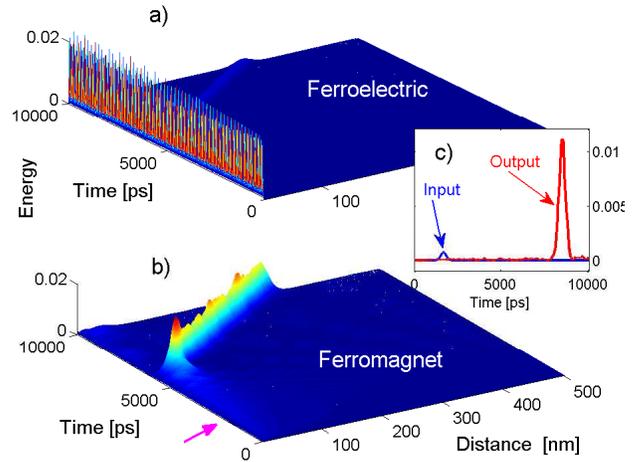}
\caption{(Color online) Results of numerical simulations of the
amplification of a magnetic signal in the multiferroic chain (a). The
space-time evolution of excitations in the ferroelectric part. (b) The
amplified propagation of the magnetic soliton in the ferromagnetic part.
The arrow indicates the moment of the injection of the magnetic
signal. Inset (c) presents the time dependence of the energy of the input
magnetic signal at the FM edge, $n=0$, while the energy of output signal is
measured at site $n=200$.}
\label{fig_2}
\end{figure}

Next, let us envisage the possibility of amplifying  magnetic
pulses: We apply a continuous electric signal with the frequency
$\omega _{s}$ which is slightly below the FM band boundary, $\omega
_{M}(0)$, keeping FM moments  pinned at the edge. In this setting,
and for small driving amplitudes, no energy is transmitted through
the chain. Both FM and FE modes are evanescent and described by the
solutions (\ref{6}). A propagating FM soliton  emerges only if the
electric-field amplitude attains the band gap transmission threshold
\cite{leon,rk}: This happens if the amplitude is large enough so
that a solution of the nonlinear dispersion relation (\ref{7m}) for
real wave number $k$ exits. Then, if one keeps the amplitude of the
electric field just slightly below this band gap threshold, a
small-amplitude in-phase magnetic signal coupled to FM chain allows
to pass the threshold. This gives rise to a large-amplitude FM
soliton propagating through the multiferroic, see Fig. 3. In this
way, one can realize an amplification of the magnetic pulses by
electric field. It may happen that almost all the energy of the
electric field will be transferred to the ferromagnet chain, and the
corresponding amplification rate may achieve values as much as $\sim
100$. For instance, in the simulations presented in Fig. 3 we choose
the driving frequency and the amplitude of the electric field
$\omega _{s}=0.229$ (that is $\approx30$ GHz in real units) and
$\left( p_{0}\right) _{\max }=0.414$, while the FM signal amplitude
is $\left( s_{0}^{x}\right) _{max}=0.02$.

In this paper we do not address dissipation effects which, in
principle, could be taken into account by introducing the
conventional Landau-Lifshitz-Gilbert damping term in the magnetic part
of the evolution equations (\ref{3}), as well as damping terms in the
electric part. Here, we assume that dissipation has no qualitative effects for
the considered length and  on the
time scales comparable with magnetic/electric signal transmission
(that is $10^{-9}$sec) and do not consider thus the respective terms in the
evolution equations.

Concluding, an electro-magnon soliton theory is developed and the
results are applied for  electric field-induced magnetic
soliton generation. A proper
choice of pump electric field parameters enables
 an amplification of magnetic signals. In the amplifying regime
the total (pump+signal) amplitude overcomes the band-gap
transmission threshold and the energy of electric field is
completely transferred to the magnetic soliton. As we have shown
above, substantial  (more than $100$ times) amplification of the magnetic
input/output signals could be realized.

\emph{Acknowledgements}.- L.Ch. and J.B.
are supported by DFG through SFB 762.             R.Kh. is
supported by DAAD fellowship and grant No 30/12 from SRNSF. The work
of B.A.M. was supported, in part, by the German-Israel Foundation
through grant No. I-1024-2.7/2009.

\end{document}